\documentclass[aps,prb,twocolumn,nofootinbib,showpacs,amsmath,amssymb]{revtex4-2}

\usepackage{bm}
\usepackage{physics}
\usepackage{graphicx}
\usepackage{mathtools}

\begin{document}

\title{Superconductivity in Jellium Model Revisited}

\author{Michael V. Sadovskii}
\footnote{E-mail: sadovski@iep.uran.ru}

\affiliation{Institute for Electrophysics, Russian Academy of Sciences,
Ural Branch\\
Amundsen str. 106, Ekaterinburg 620016, Russia}



\begin{abstract}

We reanalyze superconductivity in jellium model within the dielectric formalism 
developed by Kirzhnits, Maksimov, and Khomskii (KMK), which is probably the
most reliable approach to this model.
The linearized KMK integral equation for superconducting transition temperature
is analyzed analytically and solved numerically by direct diagonalization to
obtain $T_c$ dependence on Wigner -- Seitz radius $r_s$.
As a first systematic extension beyond random -- phase approximation (RPA), 
static Hubbard local -- field corrections are incorporated into the dielectric 
function. Our results in general narrow the interval of possible $T_c$ values
in comparison with widely scattered results of some of the previous works.
For the case of metallic hydrogen our calculations show
$T_c(r_s)$ dependence with characteristic dome with maximum of
$T_c$ at $r_s\sim$ 9.0 of the order of some fractions of Kelvin 
only, despite the naive
expectations based on the high values of pairing Boson frequency.
This is due to the weak -- coupling regime of
superconductivity in jellium model at all densities. 

\end{abstract}

\maketitle

\section{Introduction}
\label{sec:intro}

Jellium model is the simple (plasma) model of a metal, which consists of
electrically neutral (structureless) gas of electrons and ions interacting via
Coulomb repulsion. With respect to superconductivity it was popularized by
De Gennes \cite{DG} and since then it was used in dozens of papers, especially
in the context of high -- temperature superconductivity \cite{GK}. Actually
this simple model demonstrates a number of remarkable properties, like the
natural emergence of longitudinal oscillations (phonons) due to screening and
superconductivity in purely Coulomb system, so that it can be considered as a
simplest microscopic model of a metal with emergent properties, characteristic
of real metals.

In the following we shall consider only the case of the gas of electrons and
protons as a possible model for metallic hydrogen, which is the actively
studied topic today since the prediction of high -- temperature in it as well
as in different hydrides and also in astrophysical objects like white dwarfs and
pulsars or Jupiter \cite{GK1,K1,A1,A2}. The optimism with respect to the
possible high values of $T_c$ in these systems was essentially based on rather
high values of ion (proton) plasma frequency, playing the role of the phonon
coupling cut-off in BCS -- like approach \cite{GK}. The experimental discovery
of high -- temperature superconductivity in hydrides \cite{Er,Hem} has greatly
stimulated also the theoretical studies in this field.

Strangely enough the estimates of $T_c$ in all of the cited theoretical papers
varied very much, and were only qualitative. There were apparently no attempts
to analyze $T_c$ of the jellium model more or less rigorously until recently.
The first such attempt was undertaken in an interesting paper by van der Marel
and Berthod \cite{VdM}, who analyzed the problem both analytically and
numerically solving BCS -- like equations with electron -- ion
interaction described in general form as the dynamically screened Coulomb
interaction:
\begin{equation}
V(q,\omega)=\frac{4\pi e^2}{q^2\varepsilon(q,\omega)}.
\label{Coul_scr}
\end{equation}
where all physics is actually contained the general dielectric function
$\varepsilon(q,\omega)$ which was taken in the simplest RPA -- approximation
typically used to describe jellium. Rather unexpectedly their numerical
calculations produced rather low values of $T_c$ at any densities, with maximum
values not exceeding 30 K. More so, their analytical analysis of Eliashberg
parameters for the jellium model produced the value of $T_c$ not exceeding
25 mK! For some unclear reason the authors has not discussed this drastic
inconsistency of these values at all, though they rightly noticed that these
unexpected results were due to a weak -- coupling superconductivity in
hydrogen jellium.

The present paper returns to the analysis of superconductivity of the jellium
model of metallic hydrogen. In our opinion the direct use of (\ref{Coul_scr}) as
a kernel of BCS gap equation, as was in fact done in Ref. \cite{VdM} is rather
inconsistent.
The problem of the use of dielectric formalism in superconductivity
theory was analyzed more than 50 years ago in a remarkable paper by Kirzhnits,
Maksimov and Khomskii (KMK) \cite{KMK}, where BCS weak -- coupling theory was
especially reformulated for the use of general interaction of form of
(\ref{Coul_scr}) with the aim of analyzing most general pairing interactions
in solids. At a time it was seen even as an alternative to
Migdal -- Eliashberg -- McMillan  theory of superconductivity (for the review
of this theory see \cite{Scal,All,Gor}) developed for the case of
electron -- phonon mechanism of Cooper pairing, which is now state of the art
approach in microscopic calculations in most of real superconductors (including
hydrides). Unfortunately, up to now the KMK formalism has not been generalized
to strong coupling superconductors and to our knowledge was never used in the
studies of real materials. However it is still conceptually important in the
studies of non -- phonon mechanisms of pairing and is ideally fit to analyze
model interactions like those in the jellium model. The KMK dielectric formalism
provides a natural framework in which retardation and Coulomb screening are
treated on equal footing as it uses the dynamically screened Coulomb interaction
(\ref{Coul_scr}) directly in equations determining $T_c$. 
In Appendix \ref{sec:KMK_derivation} we present a brief derivation of KMK formalism
with the special emphasis on jellium model.

Within this framework:
\begin{enumerate}
\item retardation emerges dynamically,
\item Coulomb repulsion and attraction are treated simultaneously,
\item no phenomenological pseudopotential is required,
\item the superconducting gap function becomes strongly energy dependent.
\end{enumerate}
In the present work we use this classic approach to analyze $T_c$ in the
jellium model of metallic hydrogen with the aim to settle the remaining
theoretical inconsistencies of these model and extending this analysis outside
the region of weak correlations.

\section{Dielectric Function and Collective Modes}
\label{sec:jellium}

We adopt the standard form of dielectric function in jellium mode \cite{DG}:
\begin{equation}
\varepsilon(q,\omega)=1+\frac{k_{TF}^2}{q^2}-\frac{\omega_0^2}{\omega^2}.
\label{jell_eps}
\end{equation}
Here the electronic part is just the simplest static Thomas -- Fermi screening
derived within RPA approximation.
Throughout this work we use atomic units: $\hbar=e=m=$1 and consider only 
hydrogen ions (anticipating application to metallic hydrogen), so that
the ionic plasma frequency is given by:
\begin{equation}
\omega_0=\sqrt{\frac{3}{Mr_s^3}},
\label{ion_plasma}
\end{equation}
where the ratio of proton and electron mass is:
\begin{equation}
M=\frac{m_p}{m}\approx1836.
\label{p_mass}
\end{equation}
The list of other parameters fo electronic subsystem is given in Appendix
\ref{sec:numerical}.

The collective ionic (phonon) mode is determined by
\begin{equation}
\varepsilon(q,\Omega_q)=0.
\label{eps_zero}
\end{equation}
This gives
\begin{equation}
\Omega_q^2=\frac{\omega_0^2 q^2}{q^2+k_{TF}^2},
\label{spectr_coll}
\end{equation}
or equivalently
\begin{equation}
\Omega_q=\omega_0\frac{q}{\sqrt{q^2+k_{TF}^2}}.
\label{spectr_collect}
\end{equation}
At small momentum,
\begin{equation}
q\ll k_{TF},
\label{qk}
\end{equation}
one obtains acoustic behavior:
\begin{equation}
\Omega_q\sim q.
\label{ac}
\end{equation}
At large momentum,
\begin{equation}
q\gg k_{TF},
\label{opt_lim}
\end{equation}
we obtain saturation:
\begin{equation}
\Omega_q\to \omega_0.
\label{sat}
\end{equation}
This is the main property of the jellium model -- it explicitly demonstrates
the emergence of sound waves (phonons) due to screening in this Coulomb system.
In some sense the same mechanism operates in real solids, where fundamentally
only Coulomb forces are present, while phonons appear as emergent excitations.

\begin{figure}
\includegraphics[clip=true,width=0.45\textwidth]{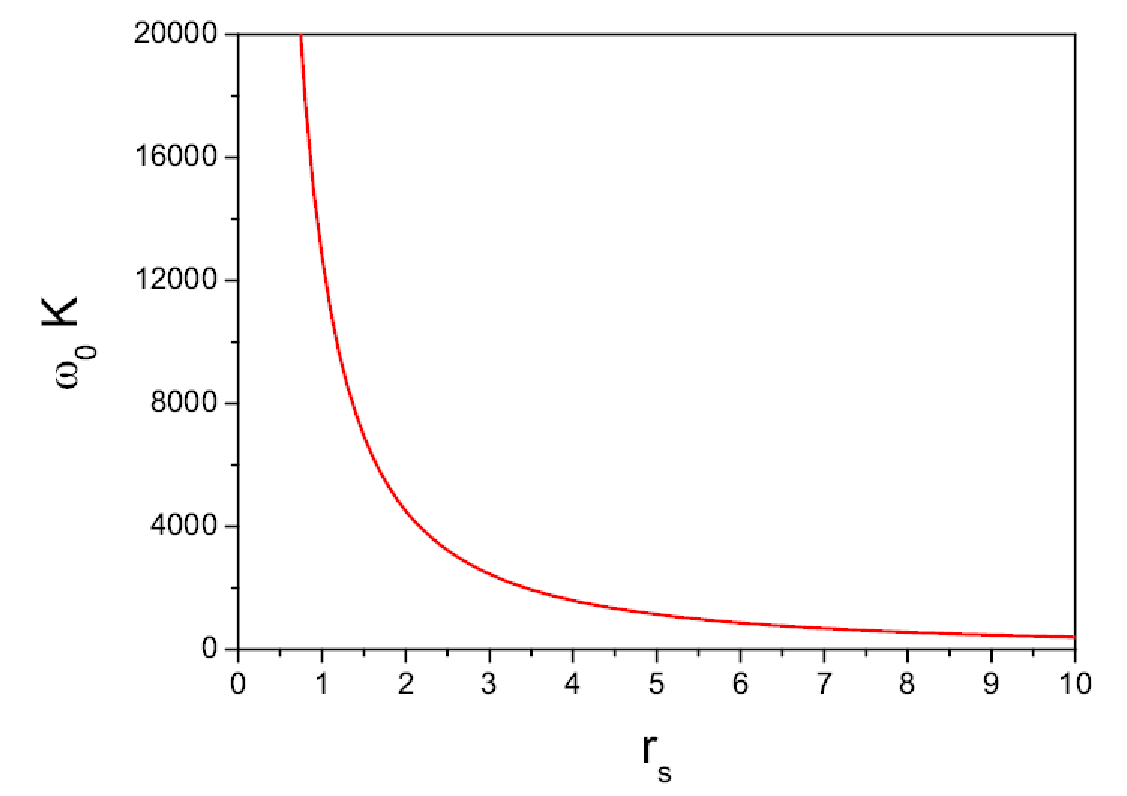}
\caption{Ion plasma frequency $\omega_0$ of jellium model as a
function of $r_s$. It is approximately the upper cut-off of phonon spectrum
in this model.}
\label{omega_0_cutoff}
\end{figure}

In Fig. \ref{omega_0_cutoff} we show the $r_s$ dependence of proton 
plasma frequency $\omega_0$ which is an effective upper cut-off of phonon
spectrum in jellium model (an analogue of Debye frequency in solids).
We can see that for typical metallic densities corresponding to $r_s\sim$ 2--3
it is pretty large.
Actually that was the driving idea of a possible high -- temperature
superconductivity in metallic hydrogen, hydrides and astrophysical objects
\cite{GK1,K1,A1,A2}.

\section{Spectral Representation}
\label{sec:spectr}

The spectral density of the inverse dielectric function which is crucial for
KMK theory is defined by
\begin{equation}
\rho(q,\Omega)=-\frac{1}{\pi}\Im\frac{1}{\varepsilon(q,\Omega+i0)}.
\label{spd_eps}
\end{equation}
Introduce
\begin{equation}
A_q=1+\frac{k_{TF}^2}{q^2}.
\label{resid}
\end{equation}
Then
\begin{equation}
\varepsilon(q,\omega)=A_q\left(1-\frac{\Omega_q^2}{\omega^2}\right).
\label{eps_qw}
\end{equation}
Therefore
\begin{equation}
\frac{1}{\varepsilon(q,\omega)}=\frac{1}{A_q}\frac{\omega^2}{\omega^2-\Omega_q^2+i0}.
\label{eps_full}
\end{equation}
Thus we immediately  obtain
\begin{equation}
\rho(q,\Omega)=\frac12\Omega_q\frac{q^2}{q^2+k_{TF}^2}\delta(\Omega-\Omega_q).
\label{spd_jellium}
\end{equation}
This will be used in all calculations which follow.

\section{Coulomb Coupling Constant}
\label{sec:Coulomb}

The statically screened Coulomb interaction is
\begin{equation}
V_C(q)=\frac{4\pi}{q^2+k_{TF}^2}.
\label{TF_Coul_static}
\end{equation}
The dimensionless Coulomb coupling is
\begin{equation}
\mu=N(0)\langle V_C(q)\rangle_{FS}.
\label{mu_const}
\end{equation}
where the averaging is over the momenta on the Fermi surface.
The transferred momentum on the Fermi surface satisfies
\begin{equation}
q^2=2k_F^2(1-\cos\theta).
\label{q2}
\end{equation}
The angular integration over the Fermi surface is performed as:
\begin{equation}
\langle f(q)\rangle_{FS}=\frac{1}{2}\int_{-1}^{1}d\cos\theta f(q(\theta)).
\label{angleav}
\end{equation}
Thus
\begin{equation}
\mu=N(0)\frac12\int_{-1}^{1}d\cos\theta\frac{4\pi}{2k_F^2(1-\cos\theta)+k_{TF}^2}.
\label{muFS}
\end{equation}
Performing the integral gives
\begin{equation}
\mu=\frac{1}{2\pi k_F}\ln\left(1+\frac{4k_F^2}{k_{TF}^2}\right).
\label{muF}
\end{equation}
Substituting the explicit expressions for $k_F$ and $k_{TF}$ we obtain
\begin{equation}
\mu(r_s)=\frac{r_s}{12.06}\ln\left(1+\frac{6.03}{r_s}\right).
\label{mu_coef}
\end{equation}

\section{KMK Pairing Kernel}
\label{sec:KMK_pairing_kernel}

The linearized superconducting ``gap'' equation
of KMK approach takes the form\footnote{Strictly speaking in KMK theory
$\Phi(\xi)$ is not a superconducting gap $\Delta(\xi)$, but $\Phi(\xi)\sim
Re \Delta(\xi)$.}
\cite{KMK}
\begin{equation}
\Phi(\xi)=-\int_{-\infty}^{\infty} d\xi' K(\xi,\xi')\frac{\tanh(\xi'/2T)}{2\xi'}
\Phi(\xi'),
\label{KMK_eq}
\end{equation}
where
\begin{equation}
K(\xi,\xi')=\mu(\xi,\xi')-2\int_0^\infty d\Omega\frac{\nu(\xi,\xi';\Omega)}
{\Omega+|\xi|+|\xi'|},
\label{kern}
\end{equation}
and $\mu(\xi,\xi')$ is Coulomb repulsion kernel which can safely be taken
as Fermi surface average:
\begin{equation}
\mu(\xi,\xi')=N(0)\langle V_C(q)\rangle_{FS}=\mu
\label{mu_kern}
\end{equation}
We also have to introduce the usual high -- energy cut-off
at $E_F$ (which is considered to be of an order of bandwidth, defining the
maximum energy in our problem) for $\xi$, $\xi'$ integration in KMK formalism.
The rest may be called an attractive contribution due to effective ``phonon'' 
exchange and written as \cite{KMK}:
\begin{equation}
K_{att}(\xi,\xi')=-2\int_0^\infty d\Omega\frac{\nu(\xi,\xi';\Omega)}
{\Omega+|\xi|+|\xi'|}.
\label{Kattr}
\end{equation}
where we also slightly simplify the general KMK equations by taking the Fermi
surface average of $\nu(\xi,\xi';\Omega)$ as:
\begin{equation}
\nu=N(0)\left<\frac{4\pi}{q^2}\rho(q,\Omega)\right>_{FS}.
\end{equation}
which is explicitly expressed via spectral density $\rho(q,\Omega)$ given by
Eq. (\ref{spd_jellium}).
Substituting this exact spectral density into Eq. (\ref{Kattr}) we get:
\small
\begin{equation}
K_{att}(\xi,\xi')=-4\pi N(0)
\left<\frac{\Omega_q}{(q^2+k_{TF}^2)(\Omega_q+|\xi|+|\xi'|)}\right>_{FS}.
\label{K_att}
\end{equation}
\normalsize
The total kernel therefore becomes
\small
\begin{align}
K(\xi,\xi')=&N(0)\left<\frac{4\pi}{q^2+k_{TF}^2}\right>_{FS}-
\nonumber\\
&-4\pi N(0)\left<\frac{\Omega_q}{(q^2+k_{TF}^2)(\Omega_q+|\xi|+|\xi'|)}
\right>_{FS},
\label{kernel}
\end{align}
\normalsize
which actually determines $T_c$ after solving the KMK integral equation
(\ref{KMK_eq}). 

\section{BCS and Eliashberg -- McMillan Approximations}
\label{sec:BCS}

At low energies $|\xi|,|\xi'|\ll \Omega_q$,
performing calculations as above one finds:
\begin{equation}
K_{att}(0,0)\approx-4\pi N(0)\left<\frac{1}{q^2+k_{TF}^2}\right>_{FS}
\equiv\lambda.
\label{K_atr}
\end{equation}
is the pairing constant of BCS theory.
Comparing Eq. (\ref{K_atr}) with Eq. (\ref{mu_const}) we immediately see that 
\begin{equation}
\lambda=\mu,
\end{equation}
which is a well known result for the jellium model \cite{VdM}.

\begin{figure}
\includegraphics[clip=true,width=0.5\textwidth]{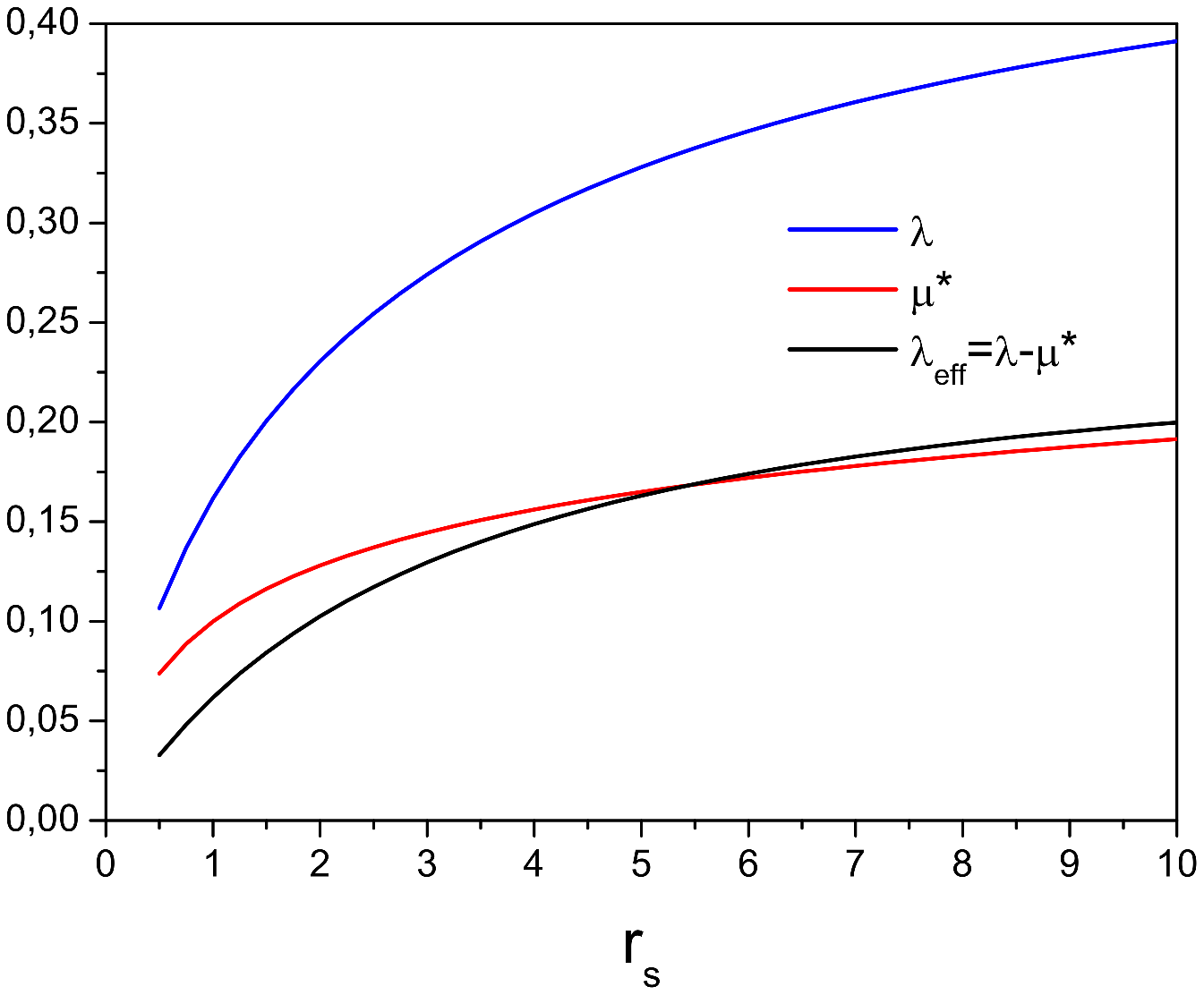}
\caption{Effective couplings $\lambda=\mu$ and pseudopotential 
$\mu^{*}$ in jellium model as functions of $r_s$, compared with effective 
pairing constant $\lambda_{eff}$.}
\label{lambda_mu}
\end{figure}

In BCS approximation it is usually assumed that
\begin{equation}
K_{att}=\lambda\theta(\omega_0-\xi)\theta(\omega_0-\xi')
\label{step}
\end{equation}
so that $\omega_0$ acts as an upper energy cut-off for electron -- phonon
interaction.

Linearized integral equation (\ref{KMK_eq}) is then approximately solved
\cite{DG} assuming:
\begin{eqnarray}
\Phi(\xi)=\Phi_{att}\theta(\omega_0-\xi) + \Phi_C\theta(E_F-\xi)
\label{Phi_BCS}
\end{eqnarray}
and reducing Eq. (\ref{KMK_eq}) to a simple system of two linear equations
for $\Phi_{att}$ and $\Phi_C$, which acquires nontrivial solution at temperatures
$T<T_c$, where $T_c$ is given by the standard BCS expression:
\begin{equation}
T_c=\frac{2\gamma}{\pi}\omega_0\exp\left(-\frac{1}{\lambda-\mu^{*}}\right)
\label{BCS}
\end{equation}
where $\gamma=1.78$ is Euler constant, and Coulomb pseudopotential $\mu^{*}$ is
given by the usual expression with Tolmachev logarithm:
\begin{equation}
\mu^{\star}=\frac{\mu}{1+\mu\ln\frac{E_F}{\omega_0}}
\label{mustar}
\end{equation}
Using here the explicit expression for $\mu(r_s)=\lambda(r_s)$ we get
$\mu^{*}(r_s)$ and $\lambda_{eff}(r_s)=\lambda(r_s)-\mu^{*}(r_s)$ for 
the effective BCS coupling from Eq. (\ref{BCS}). The appropriate $r_s$ 
dependencies of all couplings are shown in Fig. 2. 
Using now these values of coupling constants in ($\ref{BCS}$)
we obtain characteristic $T_c(r_s)$ dependence shown at the insert in Fig. 3
\cite{FPS}. The appearance of the dome is quite natural if you just look at $r_s$
dependence of parameters entering BCS expression. What is actually surprising
are the low values of $T_c$ with maximum of 3.28 K only achieved at $r_s=$ 7.50,
despite large values of the pre - exponential factor of $\omega_0$.
This contradicts optimistic
expectations expressed in Refs. \cite{GK,GK1,K1,A1,A2}. But explanation is quite
simple -- the coupling is always weak in jellium model with $\lambda_{eff}$
only slightly  exceeding 0.1 at for all realistic values of $r_s$ as can be seen
from Fig. 2.

The situation becomes even worse if we use the same coupling parameters 
in McMillan's expression for $T_c$ \cite{VdM}:
\begin{equation}
T_c=\frac{\omega_0}{1.45}\exp\left\{-\frac{1.04(1+\lambda)}{\lambda-
\mu^*(1+0.62\lambda)}\right\}
\label{McM}
\end{equation}
giving even lower values of $T_c$ in milli -- Kelvin range, as shown by the blue 
line in Fig. 3. 

Below we shall numerically solve KMK integral equation (\ref{KMK_eq}) to obtain
more rigorous estimate of $T_c$ in jellium model.

\section{Numerical Solution of the KMK Equation}
\label{sec:numerics}

Standard BCS -- like  approximation neglects non -- trivial dependence of 
kernel $K_{att}(\xi,\xi')$ (\ref{Kattr}) on $\xi$ and $\xi'$
by  putting them on the Fermi surface
and introducing the sharp integral cut-off at $\omega_0$
in (\ref{KMK_eq}). Actually this cut-off is rather smooth and fully determined
by attractive kernel dependence on $|\xi|$ and $|\xi'|$. At large energies,
\begin{equation}
|\xi|+|\xi'|\gg \Omega_q,
\label{xigg}
\end{equation}
we obtain
\begin{equation}
K_{att}(\xi|,\xi')\sim-\frac{\Omega_q}{|\xi|+|\xi'|}.
\end{equation}
Thus the attractive interaction continuously disappears while Coulomb repulsion
remains (up to the large energies of the order of $E_F$).
The superconducting eigenfunction $\Phi(\xi)$ therefore changes sign in energy
space. At low energies $\Phi(\xi)>0$, while at high energies $\Phi(\xi)<0$.
This gives a dynamic mechanism of generating Coulomb pseudopotential.

In the following we directly solve KMK integral equation (\ref{KMK_eq})
determining $T_c$ with kernel defined in (\ref{kernel}) and determine $T_c(r_s)$
dependence numerically. The numerical implementation proceeds through the
following steps:

\subsection{Energy Grid}

A logarithmic mesh is used:
\begin{equation}
10^{-10}E_F\le \xi \le E_F.
\end{equation}
which, by the way, introduces abovementioned integration cut-off in
Eq. (\ref{KMK_eq}).

\subsection{Hermitian Symmetrization}

Define
\begin{equation}
w_i=\frac{\tanh(\xi_i/2T)}{2\xi_i}\Delta \xi_i.
\label{wi}
\end{equation}
where 
\begin{equation}
\Delta\xi_i
=
\frac12
(\xi_{i+1}-\xi_{i-1}).
\label{Deli}
\end{equation}

The kernel is symmetrized according to
\begin{equation}
\widetilde K_{ij}=\sqrt{w_i}K_{ij}\sqrt{w_j}.
\label{sym_ker}
\end{equation}
This produces a Hermitian matrix with real eigenvalues.

\subsection{Transition Temperature}

The superconducting transition temperature is determined from
\begin{equation}
\lambda_{max}(T_c)=1.
\label{eigen}
\end{equation}
The largest eigenvalue is obtained by direct diagonalization.

More details on numerical implementation can be found in 
Appendix \ref{sec:numerical}.

\begin{figure}
\includegraphics[clip=true,width=0.52\textwidth]{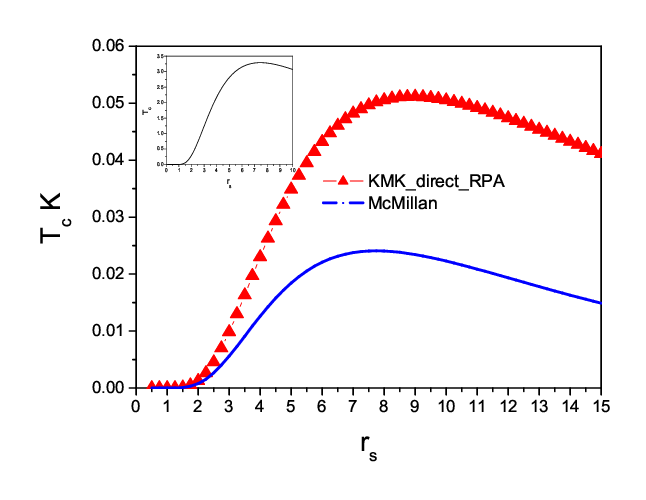}
\caption{Critical temperature of superconducting transition $T_c$ in
jellium model as a function of $r_s$. Red triangles -- direct numerical
solution of KMK integral equation, blue line -- McMillan formula for $T_c$. 
At the insert: BCS approximation. }
\label{Tc(rs)}
\end{figure}

\section{Numerical Results within RPA}
\label{sec:rpa_results}

First we present the results for the standard jellium model with
$\varepsilon(q,\omega)$ defined by Eq. (\ref{jell_eps}) and obtained within RPA.
The resulting calculations exhibit a dome-like dependence on the
Wigner-Seitz radius shown in Fig. 3 (red triangles).
At small $r_s$ the electron gas is weakly interacting and the
effective attraction remains extremely small.
Increasing $r_s$ enhances the importance of retardation effects and
increases the pairing strength.
For sufficiently large $r_s$, however, screening weakens and the
assumptions underlying the RPA dielectric function become
progressively less reliable.
The maximum of $T_c\approx$ 0.051 K is achieved at $r_s=$ 9.0, which is
relatively close to maximim $T_c\approx$ 0.024 K obtained from McMillan
expression at $r_s=$ 7.75.

The superconducting eigenfunction exhibits the characteristic
sign -- changing structure predicted by the KMK formalism, with a
positive low-energy sector and a negative high -- energy tail
(see Appendix \ref{sec:eigen}).

Among other features this behavior provides a direct microscopic realization of
Coulomb pseudopotential physics of the usual BCS theory without an artificial
cut-off at $\omega_0$.

The obtained maximal $T_c$ is orders of magnitude lower that that obtained
by numerical analysis in \cite{VdM}. Van der Marel -- Berthod approach
effectively inserts the dynamically screened Coulomb interaction into BCS gap 
equation using the substitution $\omega\to \xi-\xi'$. This yields a kernel
depending on frequency transfer $V(\xi-\xi')$. Such type of approach was
criticized by KMK \cite{KMK} many years ago. They stressed that the genuine
pairing kernel is derived from the spectral density of the inverse dielectric
function (\ref{spd_eps}), which depends on positive $|\xi|$ and $|\xi'|$
separately. KMK kernel is smooth, real and automatically retarded. 
In a sense it justifies the standard BCS -- like approach to
superconductivity and is much more reliable for calculations of $T_c$.

Surprisingly our $T_c$ values are also much lower than those obtained from
BCS expression (\ref{BCS}) and are closer to that following from McMillan's
(\ref{McM}). McMillan's transition temperature is strongly suppressed in
comparison to that of BCS mainly due to mass renormalization term term
$1+\lambda$ in the exponent of (\ref{McM}), which is due to strong -- coupling
effects of Eliashberg theory. KMK theory is usually considered a weak --
coupling approximation neglecting mass renormalization effects. We believe
that the observed $T_c$ suppression in comparison with BCS is due to
combination of nontrivial $\xi$ dependence of Eigenfuctions $\Phi(\xi)$
of Eq. (\ref{KMK_eq}) and those of the kernel $K(\xi,\xi')$ in the same
equation. Eigenfuctions of Eq. (\ref{KMK_eq}) are discussed in Appendix
\ref{sec:eigen}.

The main conclusion is that direct solution of the linear integral equation
(\ref{KMK_eq}) produces the values of $T_s$ of the order of fractions of Kelvins
for all reasonable values of $r_s$.

\section{Beyond RPA: Hubbard Local-Field Corrections}
\label{sec:hubbard}

Strictly speaking all expressions for jellium model were derived within RPA
perturbation theory approximation for electron gas, which is valid only for
$r_s$<1. The RPA neglects all exchange -- correlation effects in the electronic
response, which become important in the region of $r_s$>1. Thus all the results
given above for this region are only of qualitative nature.

There is a long history of theoretical attempts to take correlation effects
into account in the theory of interacting electron gas. Here we shall only
consider a first extension beyond RPA incorporating the so called static Hubbard
local-field corrections, which approximately decribes exchange correlation
``hole'', forming around each electron, which partly suppresses screening
at small distances \cite{Hubb}. It is commonly believed that Hubbard
local-field correction extends the validity of RPA to the region of
$r_s\sim$3--5.
This local-field correlation factor was chosen by Hubbard in the following form

\begin{equation}
G_H(q)
=
\frac12
\frac{q^2}
     {q^2+k_F^2}.
\label{GH}
\end{equation}

This expression satisfies the physically desirable limits

\begin{align}
G_H(q)
&\rightarrow 0,
\qquad
q\ll k_F,
\\
G_H(q)
&\rightarrow \frac12,
\qquad
q\gg k_F.
\end{align}

Then the dielectric function becomes

\begin{equation}
\varepsilon_H(q,\omega)
=
1+
\frac{k_{TF}^2}
     {q^2}
\left[
1-G_H(q)
\right]
-
\frac{\omega_0^2}
     {\omega^2}.
\end{equation}

The corresponding screening denominator is

\begin{equation}
D_H(q)
=
q^2
+
k_{TF}^2
\left[
1-G_H(q)
\right].
\label{DH}
\end{equation}

The collective mode in jellium is modified accordingly:

\begin{equation}
\Omega_H(q)
=
\omega_0
\frac{q}
     {\sqrt{D_H(q)}}.
\label{OH}
\end{equation}

The Coulomb contribution becomes now

\begin{equation}
K_C^{(H)}
=
N(0)
\left<
\frac{4\pi}
     {D_H(q)}
\right>.
\end{equation}

Similarly, the attractive contribution becomes

\begin{equation}
K_{\rm att}^{(H)}
=
-4\pi N(0)
\left<
\frac{\Omega_H(q)}
     {D_H(q)
      \left[
      \Omega_H(q)+|\xi|+|\xi'|
      \right]}
\right>.
\label{KH}
\end{equation}

\begin{figure}
\includegraphics[clip=true,width=0.52\textwidth]{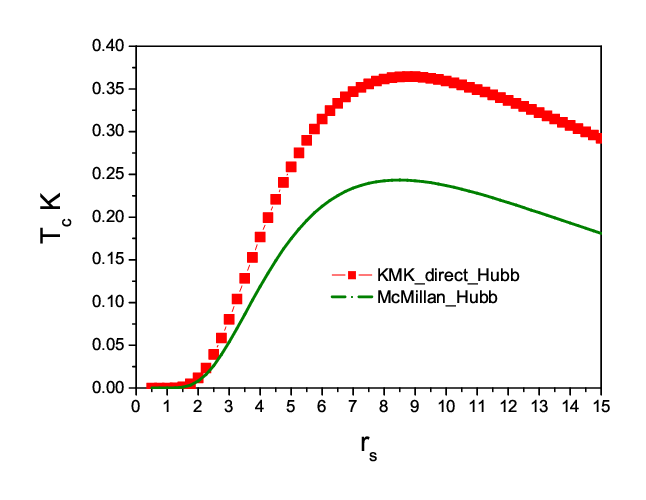}
\caption{Critical temperature of superconducting transition $T_c$ in
jellium model as a function of $r_s$ with the account of Hubbard 
local--field correction. Red squares -- direct numerical solution of KMK
integral equation, green line -- McMillan formula for $T_c$ obtained
with $\lambda=\mu$ increased due Hubbard local -- field correction.}
\label{Tc(rs)_Hubb}
\end{figure}

Equations (\ref{DH})--(\ref{KH}) constitute the central result of
the beyond-RPA extension developed in the present work.
Physically, local-field corrections act in two competing ways.
First  they reduce the effectiveness of long-range screening by
decreasing the dielectric denominator.
Second, they suppress large-momentum Coulomb scattering through the
exchange-correlation hole surrounding each electron.
The resulting influence on superconductivity is therefore highly
nontrivial.

Because the superconducting instability originates from a delicate
cancellation between repulsive and attractive contributions,
relatively small local-field corrections may produce substantial
changes in the transition temperature.

Our results shown in Fig. \ref{Tc(rs)_Hubb} demonstrate that $T_c$ is
increased several times, as compared to pure RPA case. Also in in this Figure
we show a comparison with corresponding values obtained from McMillan expression
(\ref{McM}) with recalculated couplings $\lambda=\mu$, accounting for
Hubbard local-field corrections. Actually their $r_s$ behavior is very similar
to that shown in Fig. \ref{lambda_mu}, but the absolute values are slightly
increased in comparison to RPA, leading to corresponding increase of $T_c$.
However maximum $T_c$ value remains less that 1K.

More complicated schemes accounting for correlation effects in electron gas may
be, in principle, incorporated into KMK scheme. But there are no reasons to
believe that these will lead to significant further increase of $T_c$ in
jellium model.

\section{Discussion and Conclusions}
\label{sec:concl}

We have presented a detailed derivation and numerical implementation of the KMK
dielectric formalism for superconductivity in jellium.
The exact pairing kernel was derived from the spectral representation of the
inverse dielectric function.
The superconducting instability emerges from delicate cancellation between
Coulomb repulsion and retarded ionic attraction, producing the coherent
picture of microscopic origin of superconductivity in this simplest possible
model with realistic interaction.

Most part of our work was devoted to discussion of contradictions between
$T_c$ values obtained in earlier papers. We believe that our use of general
KMK dielectric function formalism allowed to settle these inconsistences.
The final conclusion is simple and rather pessimistic -- for no reasonable
set of jellium model parameters we were not able to obtain $T_c$ larger than
fractions of Kelvin.

In general $T_c$ demonstrates a typical dome dependence on $r_s$ which in a
simple BSC or Eliashberg -- McMillan approaches directly follows
from competition between $\omega_0$ dependence on $r_s$ leading to its very
large values at small $r_s$ and $r_s$ dependence of coupling constants
$\lambda=\mu$ and $\mu^*$. Direct numerical solution of KMK equations fully
confirm this picture, solving the problem for microscopic interaction which
are determined from dynamically screened Coulomb interaction with dynamic
dielectric function $\varepsilon(q,\omega)$ of the jellium model.

Apparently there is no way to change these results significantly within
simple generaliztios of jellium model, such as more correct account of
correlation (local field) effects in electron gas, which we also considered
in a simlest possible Hubbard approximation.

In general our results are in agreement with the famous arguments of Anderson
and Cohen \cite{AndCoh}, who showed long ago that in any model with $\lambda\leq\mu$
it is not possible to obtain $T_c$ larger than few Kelvins, no matter how large
is an energy of collective excitations leading to pairing. Please note, that
these arguments are just inapplicable in a general case due to a possibility
of realistic systems with $\lambda>\mu$ \cite {KDM}, especially in the strong
coupling limit of Eliashberg theory \cite{Sad_OFN}.

The interest to superconductivity in jellium was revived in Ref. \cite{VdM}
in the context of metallic hydrogen and superconductivity in hydrides. Our
results (as well as those of Ref. \cite{VdM}) clearly show that this model is
inappropriate starting point in this field, as the values of $T_c$ obtained are
very small, due to realization of weak coupling for all parameters of the model.
It is clear from the very beginning that jellium model probably describes some
kind of liquid (or even gaseous, plasma like) state of metallic hydrogen, which
has almost nothing to do with predicted solid metallic hydrogen or hydrides with
their specific crystal structures, comlicated multiple band electron spectra and
sometimes rather exotic Fermi surfaces and multiple branches in phonon spectra.
These aspects of real hydrides, as well as of possible solid metallic hydrogen
are responsible to realization of very strong electron -- phonon pairing in
these systems in a sense of Allen -- Dynes approach to Eliashberg theory 
\cite{AD}, which allows to achieve very large values of $T_c$ up to room
temperatures \cite{Sad_OFN,Sad_24}.

At the same time the toy jellium model is still very interesting as probably the
only solvable model, which can fully describe the origin of phonon -- like
excitations and superconductivity in microscopic model with only Coulomb
interactions. In real solids all spectra of collective excitations are in fact
originating from Coulomb interactions, but we are still very far from full
microscopic theory explaining the emergence of these excitations, more so from
the real microscopic theory of superconductivity. In this sense jellium model
gives us a hope that such theory can be sometimes constructed.

\subsection{Acknowledgements}

The author is grateful to E.Z. Kuchinskii for useful discussions and
help during this work.

\subsection{Codes Availability}

All codes used in these calculations are available from the author
on a resonable request.

\appendix

\section{Derivation of KMK kernel}
\label{sec:KMK_derivation}

Below we present for completeness a short and slightly simplified version of KMK
equations with special emphasis on jellium model.

The dielectric formulation of superconductivity developed by
Kirzhnits, Maksimov, and Khomskii (KMK) \cite{KMK}
expresses the pairing interaction directly through the
frequency-dependent dielectric function of the electron-ion plasma.
Unlike Eliashberg theory, where phonons are introduced explicitly,
the KMK approach starts from the exact screened Coulomb 
interaction
\begin{equation}
W(q,\omega)
=
\frac{V_C(q)}
     {\varepsilon(q,\omega)},
\label{eq:Wexact}
\end{equation}
where\footnote{For cleareness here we write down electric charge $e$ explicitly}
\begin{equation}
V_C(q)
=
\frac{4\pi e^2}{q^2}
\label{eq:VC}
\end{equation}
is the bare Coulomb interaction.

Linearized KMK equation is derived \cite{KMK}
using the spectral representation of
$\varepsilon^{-1}(q,\omega)$.
Causality implies that
is analytic in the upper half-plane of complex $\omega$.
Therefore it admits the Kramers-Kronig representation
\begin{equation}
\varepsilon^{-1}(q,\omega)
=
1+
\int_0^\infty d\Omega\,
\rho(q,\Omega)
\left[
\frac{1}{\Omega-\omega}
+
\frac{1}{\Omega+\omega}
\right],
\label{eq:spectral}
\end{equation}
and
\begin{equation}
\rho(q,\Omega)
=
-\frac1\pi
\Im
\varepsilon^{-1}(q,\Omega+i0)
\label{eq:rho}
\end{equation}
is the spectral density of the inverse dielectric function.

Substituting Eq.~(\ref{eq:spectral}) into Eq.~(\ref{eq:Wexact})
gives
\begin{align}
W(q,\omega)
&=
V_C(q)
\nonumber\\
&+
V_C(q)
\int_0^\infty d\Omega\,
\rho(q,\Omega)
\left[
\frac{1}{\Omega-\omega}
+
\frac{1}{\Omega+\omega}
\right].
\label{eq:Wspectral}
\end{align}
The first term here is instantaneous Coulomb repulsion,
while the second contains all retarded effects.

The exact linearized Gor'kov equation in Matsubara representatiob can be
written as
\begin{align}
\Phi(k,i\omega_n)
=
-T
\sum_m
\int
\frac{d^3k'}{(2\pi)^3}
W(q,i\omega_n-i\omega_m)
\nonumber\\
\times
G(k',i\omega_m)
G(-k',-i\omega_m)
\Phi(k',i\omega_m).
\label{eq:cooper}
\end{align}
Near $T_c$ the normal-state Green functions are
\begin{equation}
G(k,i\omega_n)
=
\frac1{i\omega_n-\xi_k}.
\end{equation}

After rather long technical manipulations involving Matsubara frequency
summation KMK \cite{KMK} obtained the following one-dimensional linear integral
equation, where $\Phi(\xi)$ is aleardy a function of a single variable $\xi$:
\begin{equation}
\Phi(\xi)=-\int_0^\infty d\xi'\,\frac{\tanh(\xi'/2T)}{\xi'}
K(\xi,\xi')\, \Phi(\xi').
\label{eq:KMK}
\end{equation}
Here the KMK kernel takes the form:
\begin{equation}
K(\xi,\xi') = \mu(\xi,\xi') - 2\int_0^\infty d\Omega \,
\nu(\xi,\xi';\Omega) \frac1{\Omega+\xi+\xi'}.
\label{eq:KMK_kernel}
\end{equation}
The spectral function entering pairing is
\begin{equation}
\nu(\xi,\xi';\Omega) = N(0)\left<V_C(q)\rho(q,\Omega)\right>_{FS}.
\label{eq:alpha}
\end{equation}
The notation $\langle \cdots\rangle_{FS}$ denotes Fermi-surface averaging
as defined in the main part of our paper above.

For the electron-ion plasma we can take
\begin{equation}
\varepsilon(q,\omega)
=
\varepsilon_e(q,\omega)
-
\frac{\omega_{i}^2}{\omega^2}.
\label{eq:epsfull}
\end{equation}
The electronic dielectric function is approximated
by its static limit,
\begin{equation}
\varepsilon_e(q,\omega)
\rightarrow
\varepsilon_e(q,0).
\end{equation}
Using Thomas-Fermi approximation,
\begin{equation}
\varepsilon_e(q)
=
1+\frac{k_{TF}^2}{q^2}.
\end{equation}
Thus
\begin{equation}
\varepsilon(q,\omega)
=
1+\frac{k_{TF}^2}{q^2}
-
\frac{\omega_{i}^2}{\omega^2}.
\label{eq:epsRPA}
\end{equation}
To shorten expressions we introduce now
\begin{equation}
D(q)
=
q^2+k_{TF}^2.
\label{eq:D}
\end{equation}
Then
\begin{equation}
\varepsilon(q,\omega)
=
\frac{D(q)}{q^2}
-
\frac{\omega_{i}^2}{\omega^2}.
\end{equation}
Inverting Eq.~(\ref{eq:epsRPA}) gives
\begin{equation}
\varepsilon^{-1}(q,\omega)
=
\frac{q^2\omega^2}
     {D(q)\omega^2-q^2\omega_{i}^2}.
\end{equation}
The collective mode of jellium model can be written as:
\begin{equation}
\Omega_q^2
=
\omega_{i}^2
\frac{q^2}{D(q)}.
\label{eq:Omegaq}
\end{equation}
Then
\begin{equation}
\varepsilon^{-1}(q,\omega)
=
\frac{q^2}{D(q)}
\frac{\omega^2}
     {\omega^2-\Omega_q^2}.
\label{eq:epsinverse}
\end{equation}
Using
\begin{equation}
\frac{\omega^2}
     {\omega^2-\Omega_q^2}
=
1+
\frac{\Omega_q^2}
     {\omega^2-\Omega_q^2},
\end{equation}
one obtains
\begin{equation}
\varepsilon^{-1}
=
\frac{q^2}{D(q)}
+
\frac{q^2\Omega_q^2}
     {D(q)}
\frac1{\omega^2-\Omega_q^2}.
\label{eq:inverseps}
\end{equation}
Then one one finds:
\begin{equation}
\rho(q,\Omega)
=
\frac{q^2\Omega_q}
     {2D(q)}
\delta(\Omega-\Omega_q).
\label{eq:rhofinal}
\end{equation}
All spectral weight here resides in the ionic collective mode.

Substituting Eq.~(\ref{eq:rhofinal}) into Eq.~(\ref{eq:alpha}):
\begin{align}
V_C(q)\rho(q,\Omega)
&=
\frac{4\pi e^2}{q^2}
\,
\frac{q^2\Omega_q}
     {D(q)}
\delta(\Omega-\Omega_q)
\nonumber\\
&=
\frac{4\pi e^2\Omega_q}
     {D(q)}
\delta(\Omega-\Omega_q).
\end{align}
Performing the $\Omega$ integration gives
\begin{equation}
K_{attr}
= - N(0)
\left<\frac{2\pi e^2\Omega_q}{D(q)}
\frac1{\Omega_q+\xi+\xi'}
\right>_{FS}.
\label{eq:Kattr}
\end{equation}
This is precisely the attractive contribution used in practical KMK calculations
in the main part of our text.

Using (\ref{eq:inverseps}) in (\ref{eq:Wexact}) we can write the total interaction
as:
\begin{equation}
W(q,\omega)=\frac{4\pi e^2}{q^2+k^2_{TF}}+\frac{4\pi e^2}{q^2+k^2_{TF}}
\frac{\Omega_q^2}{\omega^2-\Omega^2_q}.
\label{eq:Weff}
\end{equation}
The second term here in KMK equation is transformed into attrative kernel
(\ref{eq:Kattr}), while the first term represents the remaining static Coulomb
repulsion:
\begin{equation}
W(q,0)=\frac{4\pi e^2}{q^2+k_{TF}^2}.
\end{equation}
The corresponding dimensionless Coulomb potential is defined as usual
\begin{equation}
\mu=N(0)\left<\frac{4\pi e^2}{q^2+k_{TF}^2}
\right>_{FS}.
\label{eq:mu}
\end{equation}
and was extensively used above.

The major simplfication of our analysis here in comparison with \cite{KMK}
is the reduction of all interactions to their Fermi surface averages. 
KMK formulation is more general with explicit ``off-shell'' expressions.
We believe that for jellium model the use of these expressions is
excessive, though  all calculations can also be performed, in principle, 
also for this general case.

\section{Main parameters and Numerical Procedure}
\label{sec:numerical}

We consider the homogeneous electron gas with density
\begin{equation}
n=\frac{3}{4\pi r_s^3},
\end{equation}
where $r_s$ is the Wigner-Seitz radius measured in atomic units.
Throughout this work we use Hartree atomic units:
\begin{equation}
\hbar=e=m=1.
\end{equation}
so that energies are measured in units of Hartree energy $Hr=mc^2\alpha^2$=
315775.0 K $\approx$ 27 eV ($OD\alpha=e^2/\hbar c$ is the fine structure constant). All
lenghts are measured in the values of Bohr radius: $a_0=\hbar/mc\alpha$
and $r_s$ is just dimensionless.
The Fermi momentum is
\begin{equation}
k_F=(3\pi^2 n)^{1/3}=\frac{(9\pi/4)^{1/3}}{r_s}.
\end{equation}
The free-electron dispersion is
\begin{equation}
\xi_k=\frac{k^2-k_F^2}{2}.
\end{equation}
The Fermi energy becomes
\begin{equation}
E_F=\frac{k_F^2}{2}.
\end{equation}
The density of states per spin projection at the Fermi level is
\begin{equation}
N(0)=\frac{k_F}{2\pi^2}.
\end{equation}
The Thomas-Fermi momentum is
\begin{equation}
k_{TF}^2=\frac{4k_F}{\pi}.
\end{equation}

The linearized superconducting ``gap''  equation is written as
\begin{equation}
\Phi(\xi)
=
-
\int_{-\infty}^\infty
d\xi'
K(\xi,\xi')
\frac{\tanh(\xi'/2T)}
     {2\xi'}
\Phi(\xi').
\label{KMK_lin}
\end{equation}
where the kernel $K(\xi,\xi')$ was defined in (\ref{kernel}).

The transferred momentum on the Fermi surface satisfies
\begin{equation}
q^2=2k_F^2(1-\cos\theta).
\end{equation}
The angular integration over the Fermi surface is performed directly using:
\begin{equation}
\langle f(q)\rangle_{FS}=\frac{1}{2}\int_{-1}^{1}d\cos\theta f(q(\theta))
\end{equation}
Energy discretization is introduced giving

\begin{equation}
\Phi_i
=
\sum_j
M_{ij}
\Phi_j.
\end{equation}
The matrix elements are

\begin{equation}
M_{ij}
=
-
K_{ij}
\frac{\tanh(\xi_j/2T)}
     {2\xi_j}
\Delta\xi_j.
\end{equation}
where
\begin{equation}
\Delta\xi_i
=
\frac12
(\xi_{i+1}-\xi_{i-1}).
\end{equation}
To obtain a Hermitian eigenvalue problem one introduces
\begin{equation}
w_i
=
\frac{\tanh(\xi_i/2T)}
     {2\xi_i}
\Delta\xi_i,
\end{equation}
and defines
\begin{equation}
\widetilde K_{ij}
=
\sqrt{w_i}
K_{ij}
\sqrt{w_j}.
\end{equation}
The superconducting instability occurs when the maximum
eigenvalue of Eq. (\ref{KMK_lin}).
\begin{equation}
\lambda_{\max}(T_c)=1.
\end{equation}
This criterion forms the basis of the numerical calculations
presented below.

The numerical solution of the KMK equation presents a nontrivial
challenge because the thermal factor

\begin{equation}
W(\xi,T)
=
\frac{\tanh(\xi/2T)}
     {2\xi}
\end{equation}
contains the Cooper singularity

\begin{equation}
W(\xi,T)
\sim
\frac{1}{2\xi},
\qquad
\xi\rightarrow 0.
\end{equation}
A uniform energy mesh would therefore require an impractically large
number of points.
To resolve both infrared and ultraviolet scales efficiently, we
introduce a logarithmic mesh
\begin{equation}
\xi_i
=
\xi_{\min}
\left(
\frac{\xi_{\max}}
     {\xi_{\min}}
\right)^{i/(N-1)},
\qquad
i=0,\ldots,N-1.
\end{equation}
In the calculations reported here we typically used

\begin{align}
\xi_{\min}
&=
10^{-10}E_F,
\\
\xi_{\max}
&=
E_F
\end{align}
with
\begin{equation}
N=200-300.
\end{equation}
This choice provides accurate resolution of both the Cooper
logarithm and the high-energy Coulomb tail and corresponds to sharp cut-off at
$E_F$. Actually our numerics has shown that the value of $T_c$ is very slowly
(logaritmically) increasing as we increase this cut-off from $E_F$ to 10$^2E_F$, 
10$^3E_F$ etc. This is unphysical growth -- introducing the smooth
cut-off beyond $E_F$ we achieve full convergence of our results for $T_c$
to the values obtained with sharp $E_F$ cut-off as we increase intgration limits
to 10$E_F$, 10$^2E_F$ and beyond.

The numerical solution proceeds as follows:

\begin{enumerate}
\item Construct logarithmic energy grid.
\item Compute angularly averaged kernel.
\item Form thermal weights.
\item Construct Hermitian matrix
\begin{equation}
\widetilde K_{ij}
=
\sqrt{w_i}
K_{ij}
\sqrt{w_j}.
\end{equation}
\item Determine largest eigenvalue.
\item Solve for
\begin{equation}
\lambda_{\max}(T_c)=1
\end{equation}
by root finding.
\end{enumerate}
This procedure yields the superconducting transition temperature
directly without introducing additional fitting parameters.

\section{Eigenfunctions}
\label{sec:eigen}

\begin{figure}
\includegraphics[clip=true,width=0.47\textwidth]{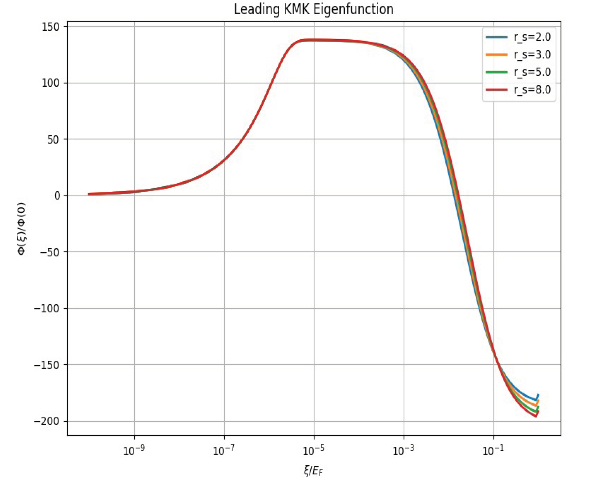}
\caption{Eigenfunctions of KMK equation $\xi$ behavior on electronic
energy scale for several typical values of $r_s$.}
\label{Eig_el}
\end{figure}

\begin{figure}
\includegraphics[clip=true,width=0.43\textwidth]{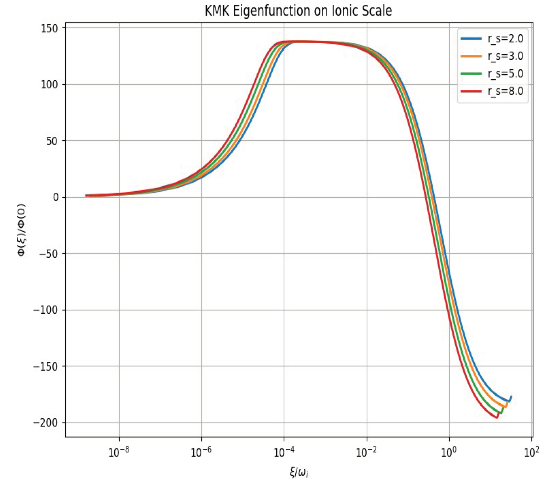}
\caption{Eigenfuctions of KMK equation $\xi$ behavior on ion energy scale
for several typical values of $r_s$.}
\label{Eig_ion}
\end{figure}

Our numerical calculations produced eigenfuntions $\Phi(\xi)$ of
Eq. ({\ref{KMK_eq}}) with $\xi$ -- dependence shown in Fig. \ref{Eig_el} and
in Fig. \ref{Eig_ion}, presenting it on electron and ion energy scales.
This behavior is almost universal with eigenfuctions changing sign at
$\xi\sim\omega_0$. This behavior reflects the
electon -- ``phonon'' nature of pairing in jellium model.

The node at $\xi\sim\omega_0$ means that below $\omega_0$ attraction 
dominates, while above $\omega_0$ Coulomb repulsion dominates.
The eigenfunction automatically separates these two regions.
This is essentially the same physics
that produces Tolmachev logarithm in BCS and Eliashberg theories.
The remarkable point is that KMK seems to generate it directly from the
eigenvalue problem without introducing a pseudopotential by hand.

Why KMK $T_c$ is so much lower than BCS? Our numerics are giving a fairly clear
answer. The naive BCS estimate assumes $\Phi(\xi)=const$ through the entire
pairing region. The actual KMK solution instead looks schematically like
$\Phi(\xi)$ positive\footnote{Actually the sign here is relative due to linear 
nature of KMK integral equation, only sign change matters.} for $\xi<\sim\omega_0$ and
$\Phi(\xi)$ negative for $\xi>\sim\omega_0$. Also for some part of low $\xi$
region $\Phi(\xi)$ is rather small.
The entire shape of $\Phi(\xi)$ matters. Our eigenfunctions are not some 
two-step functions. They are rather smooth objects extending over many decades 
in energy. Consequently the largest eigenvalue of the
kernel is just smaller than one would infer from a simple averaged coupling
picture. That naturally lowers $T_c$.


\end{document}